\documentclass[lettersize,journal]{IEEEtran}
\usepackage{amsmath,amsfonts}
\usepackage{algorithmic}
\usepackage{array}
\usepackage[caption=false,font=normalsize,labelfont=sf,textfont=sf]{subfig}
\usepackage{cite}
\usepackage{xcolor}
\usepackage{hyperref}
\usepackage{textcomp}
\usepackage{stfloats}
\usepackage{url}
\usepackage{verbatim}
\usepackage{graphicx}
\usepackage{paralist}
\def\BibTeX{{\rm B\kern-.05em{\sc i\kern-.025em b}\kern-.08em
    T\kern-.1667em\lower.7ex\hbox{E}\kern-.125emX}}
\usepackage{balance}

\begin{document}
\title{Meta Computing}
\author{Author
\thanks{xxx}
}

\markboth{Journal of \LaTeX\ Class Files,~Vol.~18, No.~9, September~2020}%
{How to Use the IEEEtran \LaTeX \ Templates}

\newcommand{\add}[1]{{\color{blue}#1}}


\author{\IEEEauthorblockN{Xiuzhen~Cheng\IEEEauthorrefmark{2}, 
Minghui~Xu\IEEEauthorrefmark{2}\IEEEauthorrefmark{1}, 
Runyu~Pan\IEEEauthorrefmark{2}, 
Dongxiao~Yu\IEEEauthorrefmark{2}, 
Chenxu~Wang\IEEEauthorrefmark{2}, 
Xue~Xiao\IEEEauthorrefmark{3},
Weifeng~Lyu\IEEEauthorrefmark{4}
}

\IEEEauthorblockA{\IEEEauthorrefmark{2}School of Computer Science and Technology, Shandong University}\\
\IEEEauthorblockA{\IEEEauthorrefmark{3}Inspur}\\
\IEEEauthorblockA{\IEEEauthorrefmark{4}State Key Laboratory of Software Development Environment, Beihang University}
\thanks{*Corresponding author: Minghui Xu (mhxu@sdu.edu.cn).}
}

\maketitle

\begin{abstract}
With the continuous improvement of information infrastructures, academia and industry have been constantly exploring new computing paradigms to fully exploit computing powers. In this paper, we propose Meta Computing, a new computing paradigm that aims to utilize all available computing resources hooked on the Internet, provide efficient, fault-tolerant, and personalized services with strong security and privacy guarantee, and virtualize the Internet as a giant computer, that is, ``Network-as-a-Computer, NaaC'', or ``Meta Computer'' for short, for any task or any person on-demand.
%
Meta computing possesses the following three functional characteristics: supporting zero-trust environments, integrating all available computing resources, and configuring meta computers on-demand. We also present an architecture for meta computers, analyze the most-influential emerging endogenous applications of meta computing, and put forward open technological challenges to realize meta computing.
\end{abstract}


\section{Introduction}
The wide adoption of the Internet has had a significant impact on the evolution of computing paradigms, leading to a shift from providing scientific computing capabilities to offering integrated network services. Such a radical change has resulted in a revolution in information systems in terms of architectures and technologies. As the new generation of information infrastructures continues to be improved, researchers have been exploring novel computing paradigms to take full advantages of the computing resources. Since the 1960s, computing paradigms have undergone significant developments, evolving from the original Client-Server model to newer ones such as Cloud Computing, IoT (Internet of Things), and Edge Computing. The early developments of the computing paradigms can be roughly divided into three categories based on the types of infrastructures: high performance back-end, personal computer front-end, and IoT terminal-end.

\begin{figure*}
    \centering
    \includegraphics[width=\textwidth]{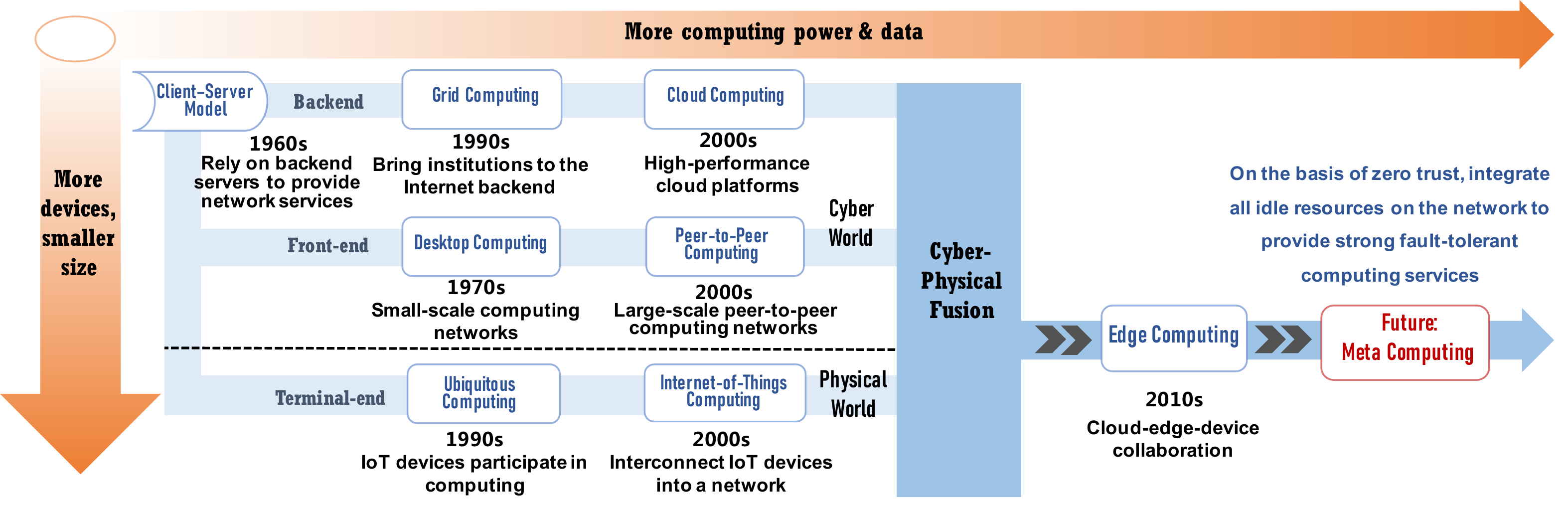}
    \caption{The development history of the computing paradigms}
    \label{fig:roadmap}
\end{figure*}

As depicted in Figure~\ref{fig:roadmap}, the increasing demands on various computing resources have driven the Client-Server model to advance along three directions, with their user groups and application domains largely non-overlapping before the end of the 20th century. 
Specifically, at the back-end the continuous development of high performance clusters has resulted in the creation of petascale supercomputing centers and cloud computing services~\cite{chellappa1997intermediaries}; at the front-end the performance of personal computers has been significantly improved with the fast-growing user base, and distributed peer-to-peer networking~\cite{oram2001peer-to-peer, barkai2001technologies} for resource sharing among desktop computers has become popular; and at the terminal-end sensing devices have become smaller and smaller in size and larger and larger in quantity, facilitated with the advances in communication technologies, gradually establishing the IoT Computing paradigm~\cite{itu2005internet}. The computing power of the high-performance back-end and personal computer front-end has established the digital infrastructure for the cyber world, while IoT terminal equipment is connected to not only the cyber world, but also the physical world to continuously collect real-time data. Furthermore, the more and more fine-grained and comprehensive perception of the physical world via IoT sensing as well as the more and more fast and effective processing of the sensed data at the cyber world drive the development of cyber-physical systems that rely on closed-loop controls for automate operations, thereby strengthening the fusion of the cyber and physical worlds and further promoting the integration of the three computing paradigm categories. Correspondingly the boundaries of their user groups and application domains are becoming more and more blurry. However, this fusion underlines the huge gap between demand and supply of computing powers, that is, the linear growth of computing resources (CPU, bandwidth, etc.) cannot meet the exponentially growing demand by the terminal devices that continuously produce an immense amount of data. This contradiction drives the emergence of Edge Computing~\cite{satyanarayanan2009case, bonomi2012fog}, whose development has been started since the early 21st century. The central theme of edge computing is to realize cloud-edge-terminal coordination, i.e., to fully utilize the idle computing resources at the network edge for data processing and shift the computing tasks to farther resources only when the closer ones are not sufficient, thereby reducing the pressures of massive data transport on the network core and expensive data processing on clouds while improving real-time performance and security. 

In recent years, researchers and developers have made a tremendous progress in edge computing technologies, and leading companies have been offering various services to exploit the computing resources at the network edge. Yet, the problem of computing power islands is getting more and more highlighted, as service providers cannot coordinate to achieve the goal of utilizing nearby available computing resources as much as possible due to the lack of trust among themselves. Additionally, the current effort focuses on the utilization of the edge resources, completely ignoring the terminal devices that are ubiquitous and have idle computing powers to share. Notice that the accumulated computing power of several dozens of micro-controllers is comparable with that of a desktop computer \cite{coremark}. Nevertheless, fully utilizing the computing resources provided by the pervasive terminal devices require complicated coordination and fault-tolerance mechanisms, as such devices are usually lightly-weighted and agile. Motivated by these observations, we propose Meta Computing, a new computing paradigm that aims to make full use of all available computing resources hooked on the Internet, break down the barriers of computing power islands, and achieve fault-tolerant network computing to accommodate low-end devices. The utmost goal of meta computing is to turn the Internet into a distributed giant computing power pool to fulfill the ever-increasing needs on computing powers. 

Meta computing is an important step pushing forward the computing paradigms to the next generation -- it is believed that a new computing paradigm comes into being every decade. The basis of meta computing lies in the recent advances in cloud, edge, and IoT computing. This paper first outlines and analyzes the development history of computing paradigms in the past 60 years, then proposes the definition, architecture, and functional goals of meta computing. This paper also presents three typical endogenous applications as examplars to demonstrate the potential of meta computing, and identifies four technical challenges for the realization of meta computing.

\section{The Development History of Computing Paradigms}
In this section, we provide a detailed overview on the development history and the state-of-the-art of mainstream computing paradigms, then explain how the limitations of the existing computing paradigms drive the emergence of meta computing  to satisfy the computing power demands by the new-generation information systems. 

\subsection{Client-Server Model}
The Client-Server model is a type of architecture proposed for distributed applications, where a server provides resources or services to the clients who make the requests. This model has a long history, with early examples dating back to IBM's OS/360 system  \cite{witt1965ibm} in 1964 to support remote requests. As a popular option, this model has been widely adopted to develop various network applications in recent years, including email, Telnet, FTP, and the Web.

In the Client-Server model, clients make a large number of requests for services from a server, which puts a lot of strain on the server, resulting in the so-called denial-of-service attacks. This has led to the developments of other computing paradigms as well as efforts to increase the power of servers to handle the high volume of requests. 
To increase capacity, server clusters have been deployed, which gradually evolve into the back-end computing facilities. Additionally, the continuously enhanced personal computer hardware provides a surplus of the network resources, making it possible for a personal computer to serve as a proxy of a server, thereby significantly decreasing the server's work load. This leads to the front-end development of computing paradigms. Finally, the advances in sensing technologies require a server to handle a huge amount of data while the networking capability of the tiny sensors is poor, driving the Client-Server model to develop along the terminal-end direction for pervasive computing. 

\subsection{Grid Computing and Cloud Computing}
In the 1990s, Grid Computing gained popularity as a technology to integrate and utilize idle computing powers of various devices to provide high-performance computing services. According to the book ``The Grid: Blueprint for a New Computing Infrastructure'' \cite{Foster1998grid} by Ian Foster and Carl Kesselman, a grid is a set of emerging technologies that uses internetworking to bring together the high-speed Internet, computers, databases, sensors, and other devices to provide scientists and mass users with more resources, functions, and services. Like the use of electric power systems, researchers expect that a user can easily access grid resources once its device is plugged into a ``socket'' on the grid.

In 1997, Chellappa introduced the concept of Cloud Computing at the INFORMS (Institute for Operations Research and the Management Sciences) annual meeting, with the primary purpose of providing a more cost-effective solution for small- and medium-sized enterprises \cite{chellappa1997intermediaries}. Unlike Grid Computing, which allows any device to join a grid, Cloud Computing intends to concentrate computing tasks into a high-performance computing cluster with a virtualized data center, so as to provide on-demand and customizable computing services for enterprises. The significant increase in the number of Internet users whose computing tasks cannot be handled locally by personal computers also drives Cloud Computing to offer high-quality services to regular users. 
In the early 21st century, Cloud Computing technology was developed rapidly, and the Internet giants such as Amazon, Google, Microsoft, IBM, and Alibaba launched their own cloud service products to capture the market shares. Ever since then, there has been a significant amount of effort in overcoming various challenges of cloud computing and continuously enhancing its service quality from both industry and academia.
 In 2022, researchers at the RISELab of the University of California, Berkeley, introduced a new concept of Sky Computing to bring together various cloud services and provide a single, unified public cloud platform~\cite{chasins2022sky}. In addition, the Compute First Networking (CFN) framework~\cite{krol2019compute} was launched in 2019 with the purpose of integrating cloud and edge computing to address the challenges of limited and unbalanced edge resources via dynamically optimizing resource allocation for high-performance computing services.

\subsection{Desktop Computing and Peer-to-Peer Computing}
The first generation personal computers in the 1970s marked the beginning of the Desktop Computing era. These computers, also called Desktop Computers, are small enough to be placed on a desk for easy use. Because of their programmability, desktop computers can complete a wide range of work and personal tasks. However, without a networking support, Desktop Computing is limited to using the computing resources offered only by the desktop computer and its input/output devices, which may not fully meet the needs of the users.

With the development of the computer network technologies and the creation of the Internet, tens of thousands of computers around the world were able to connect and collaborate. The Internet has made it possible for new services such as email and the web to be widely used. However, with the increase of the number of users, servers under the client-server model began to struggle with limited resources such as network bandwidth and CPUs, especially when the requests are bursty and unbalanced in time, making it difficult to support high-quality large-scale resource sharing. Even worse, a server may become a single point of failure that can result in service unavailability when it crashes, causing large financial losses and poor user experiences. To overcome these issues, a new computing paradigm called Peer-to-Peer Computing came into being in the early 21st century. This paradigm significantly increases the utilization of network-wide resources and alleviates the issues mentioned above. It gains popularity also due to its emphasis on equality. Peer-to-Peer Computing relies on a distributed network architecture to equally distribute tasks to each peer node, with the goal of maximizing the use of network bandwidth, storage space, idle processor cycles, and other resources at the network edge \cite{oram2001peer-to-peer}. Peer-to-Peer Computing possesses three important characteristics \cite{barkai2001technologies}: service tasks such as calculation and file sharing are performed on the network edge (e.g., via personal computers); network resources including bandwidth, storage, CPUs, and user contents are shared across the entire network; and an overlay network is created among peer nodes.

\subsection{Ubiquitous Computing and IoT Computing}
Ubiquitous Computing was first proposed in 1988 at the Xerox PARC Computer Science Laboratory by Mark Weiser. In the 1990s, as personal computers became more prevalent in everyday life, more and more work and leisure activities were moved to computers. However, the complex command-line, menu, and user interfaces made it difficult for the general public to enjoy the benefits brought by computers. Weiser believed that the best technology should be ``invisible'', or fully integrated into people's environments and daily lives, so that human beings are unaware of its existence \cite{weiser1999computer}. This is the fundamental idea behind Ubiquitous Computing. In a ubiquitous computing environment, users interact with numerous computing devices in their surroundings without knowing how the service is provided. Just like when people turn a light on and off, they don't need to know how the circuit is connected or where the electric power comes from \cite{zheng2003ubiquitous}. Ubiquitous computing is characterized by the following three main traits: ubiquity, interconnectivity, and dynamism \cite{zheng2003ubiquitous}. Ubiquity refers to the presence of a wide range of embedded devices in daily life, covering people's needs. Interconnectivity means that all computing devices are connected to a network via wired or wireless technologies and can support high-speed data transmissions. Dynamism allows mobile devices to constantly move in and out or change connection locations and quickly reconfigure themselves. With the widespread adoption of personal mobile devices such as smart phones and tablets, people are getting closer to realizing ubiquitous computing. The rapid development of the Internet of Things (IoT) technology in recent years has also brought whole society closer to ubiquitous computing.

IoT refers to a network of interconnected devices with the ability to sense, transmit, and process data. It makes use of technologies such as radio frequency identification (RFID), sensor networking, and intelligent computing to connect the physical and cyber worlds, enabling interconnections among things in the globe. The concept of IoT can be traced back to the RFID networking system proposed by MIT's Auto-ID Laboratory, which connects items to the Internet via techniques such as video recognition to enable intelligent identification and management of items. In 2005, the International Telecommunication Union (ITU) released "ITU Internet Reports 2005 - the Internet of Things" \cite{itu2005internet} at the World Summit on the Information Society (WSIS) in Tunisia, officially clarifying the concept of the Internet of Things.

IoT is characterized by comprehensive perception, reliable transmission, and intelligent processing \cite{sun2010internet}. This implies that IoT is not just a simple sensor network -- it requires intelligent information processing and utilization. IoT is the first step towards the realization of ubiquitous networking and computing. Its ultimate goal is to enable people and devices to connect to services with minimal technical restrictions via any approach, at any time, and in any location \cite{itu2009overview}.

\subsection{Edge Computing}

Facilitated with IoT technologies, a better perception and cognition of the physical world can be obtained in real-time, driving the development of Cyber-Physical Systems (CPS). As a technology, CPS can not only perceive the physical world, but also manipulate physical entities based on the perceived data, creating a closed-loop feedback control that can significantly enhance the intelligence and autonomy of a system \cite{wolf2009cyber}. 
Modern complex information systems today rely on CPS, which is closely tied to the development of IoT technologies. However, the ability of CPS relies on reliably sensing, transporting, and processing data in real-time, which requires sufficient computing power to be instantly available whenever needed. This puts a grand challenge, as CPS systems usually sit at the network end. Completely shifting the computational tasks to clouds is not always feasible due to the real-time requirement as well as security and privacy concerns. Simply enhancing the computing infrastructure cannot solve the problem too as the expanding rate of high-performance networking and computing facilities cannot keep up with the growing speed of the involved IoT device and data scale.  
To overcome this dilemma, Edge Computing was proposed, whose idea is to utilize the geographically-collocated or nearby computing resources at the  network edge as much as possible while shifting the data processing tasks to the remote clouds only when necessary.  Such a smart concept can also exploit the generally idle or under-utilized computing resources at the network edge, which are usually in abundance but are wasted most of the time. Edge computing requires the coordination among cloud, edge, and terminal layers to enable data processing by the edge servers that are as close to applications as possible, constrained by the real-time and other system performance requirements.  This drives the development of a hierarchical cloud-edge-terminal computing architecture for edge computing, to enable the fast processing and feelback control of CPS systems. 

The CDN (Content Delivery Network) proposed by Akamai in the 1990s was deemed as one of the earliest examples that embody the concept of edge computing~ \cite{dilley2002globally}. CDN achieves the goal of saving network bandwidth by pre-fetching and caching contents at the edge near the users. In 2009, Satyanarayanan \textit{et al.}~ \cite{satyanarayanan2009case} proposed a two-layer network architecture, Cloud-Cloudlet, to reduce the processing delay by setting up cloudlet nodes at the edge. This work laid the foundation for the concept of edge computing. In 2012, Bonomi et al. presented a similar concept called Fog Computing~ \cite{bonomi2012fog}, to utilize fog nodes for taking on some of the computational overhead of cloud servers. In industry, Nokia and IBM jointly introduced the RACS (Radio Applications Cloud Server) platform for 4G/LTE networks in 2013. Vodafone, Intel, Huawei, and CMU collectively announced the Open Edge Computing Initiative (OEC) in 2015 to establish cooperative laboratories for deploying cloudlet-based applications and accumulating practical operational experience. In 2017, Amazon introduced AWS Greengrass, which allows for the deployment of Amazon AWS services to edge devices. In 2020, Google introduced the GMEC (Global Mobile Edge Cloud) telecommunications-specific platform for global mobile edge cloud computing. In China, Tencent Cloud, a subsidiary of Tencent Company, introduced the TSEC (Tencent Smart Edge Connector) in 2019; Alibaba announced its ENS (Edge Node Service) and Link Edge IoT edge computing products at the same year, and open-sourced its edge computing project in 2020.

\subsection{Limitations of Existing Computing Paradigms}

\begin{figure}[!t]
	\centering
	\includegraphics[width=0.45\textwidth]{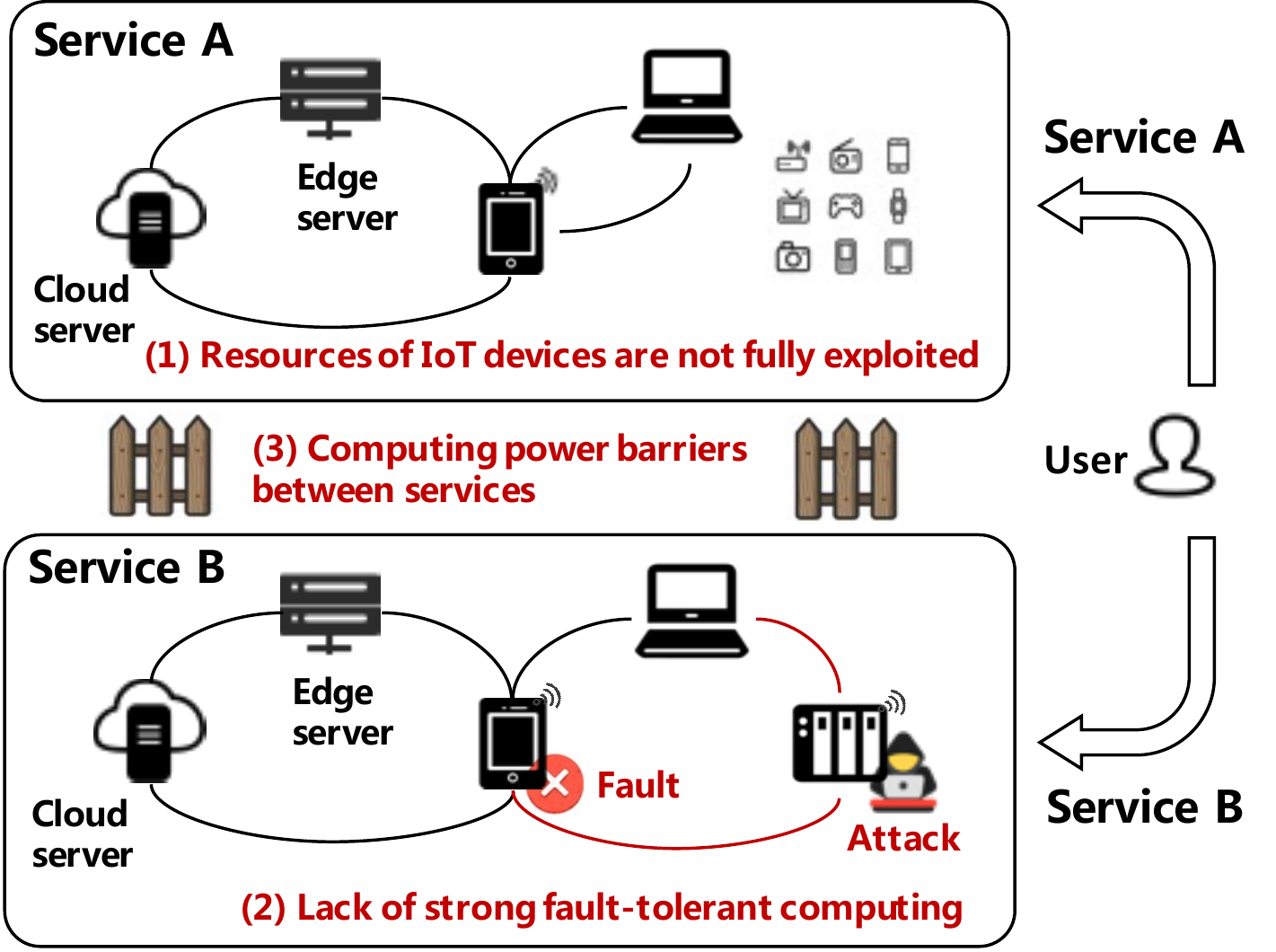}
	\caption{Problems of existing computing paradigms}
	\label{fig:problem}
\end{figure}

Nowadays, perhaps the most influential computing paradigms are edge computing and cloud computing, with both playing significant roles in the social-economic development and construction all over the world. Nevertheless, none of them is the terminator of computing paradigms, leaving a spacious room for further development.
In this subsection, we analyze the three major limitations of existing computing paradigms, which motivate the development of our novel meta computing.



\textbf{(1) Terminal-end IoT resources are not fully exploited.} 
The continuous improvement of terminal capabilities in IoT systems is driving the evolution of computing paradigms, but the potential of terminal-end computing power has not yet been fully utilized. Edge computing primarily relies on the edge-side computing power rather than taking advantage of that at the terminal-end. However, as terminal-end resources become increasingly important due to their ever-increasing quantity and quality~\cite{warneke2001smart}, they will become a significant part of a computing paradigm in future. In fact, the integrated computing power of the IoT devices in a small-scale system could be huge, and sufficiently big enough to support real-time data-intensive applications. For instance, the combined computing power of dozens of microcontrollers in a smart home is comparable to that of a personal computer \cite{coremark}; also, the microcontrollers are closer to the sensors and actuators, thus the computing latency is shorter. Therefore future computing paradigms should consider to exploit the unused or under-utilized computing resources at the terminal-end, especially the ubiquitous IoT devices, which can enable more real-time services and enhance security. 


\textbf{(2) There exist barriers to the integration of network-wide computing power.}
While significant amount of effort has been made to integrate the resources at the cloud, edge, and terminal layers, which is the major objective of edge computing,  a consolidation of a network-wide computing resources is far from being reached. Sky computing was an early effort to provide a resource integration for cloud computing, but it is constrained within clouds. In fact, edge computing service providers have made big progress to integrate their own resources in recent years; but due to the lack of mutual trusts, geographically collocated resources from different providers cannot be used together. Even within the same service domain, resource consolidation is unsatisfactory because appropriate techniques that can handle the device and protocol heterogeneity are not mature enough. Such phenomena cause the computing power islands problem, which recently has caught extensive attentions. 


\textbf{(3) There is a lack of strong fault tolerance in current computing paradigms.} 
It is undoubted that guaranteeing the correctness and accuracy of the computing results is of the first importance for all computing systems. This is a non-trivial task due to the multi-dimensional complexity brought by input, computing facilities and environment. Security and privacy attacks also affect the accuracy of the computing results and the reliability of the computing systems. Traditional computing paradigms have been dedicated to building ``perfect'' systems that rely on ``perfect'' software and hardware as well as ``perfect'' data to yield ``perfect'' results. But such an ideal situation does not exist in reality; therefore a huge amount of effort has been made for testing and vulnerability mining in the past years. Nevertheless, human beings can hardly win this endless battle, especially when the low-end fragile IoT devices and lightly-weighted protocols tailored for IoT, which are not trustworthy at all, are taken into account. This leaves fault-tolerance an indispensable property that must be possessed by future computing paradigms. In fact, fault-tolerance computing, which seeks verifiably correct results without requiring all computing components to be fault-free, has been ever-present, and the success of blockchain has significantly elevated its importance. 


In this article, we propose meta computing, a new computing paradigm to address these three issues. One can see that a deep exploitation of the resources provided by the terminal-end IoT devices and those by other service providers, and an effective consolidation of the network-wide computing resources,  can  provide a green solution to help solve the following dilemma: emerging applications desperately need more computing power while available computing powers are ubiquitous but cannot be easily used. Additionally, aggressively utilizing the available unused or under-utilized resources hooked on the Internet requires fault-tolerance computing to handle the possible faults brought by untrusted hardware/software, especially those of the low-end fragile IoT devices.

\section{Meta Computing}

\subsection{Motivation and Definition}

As mentioned earlier, computing power shortage is getting more and more severe, gradually becoming the most critical constraining factor for social and economic developments. Various techniques and mechanisms have been proposed to mitigate the pressure but none of them has been well-received as an effective and viable approach. For example, widely deploying a large number of data centers that can cover as broad geographic areas as possible, which has been adopted by a few countries, has been deemed as a costly and undesirable solution. We believe that tackling the problem from the perspective of computing paradigms and overcoming their limitations mentioned above to utilize the ubiquitous unused or under-utilized computing resources hooked on the Internet, provides a promising and green solution, which motivates our proposal of meta computing.

Formally speaking, meta computing is a new computing paradigm that aims to utilize all available computing resources hooked on the Internet, provide efficient, fault-tolerant, and personalized services with strong security and privacy guarantee, and virtualize the Internet as a giant computer, that is, ``Network-as-a-Computer, NaaC'', or ``Meta Computer'' for short, for any task or any person on-demand. This definition can be interpreted from three perspectives. First, meta computing intends to utilize all available computing resources hooked on the Internet to the maximum, to address the first two limitations of current computing paradigms. This requires innovative technologies that can effectively integrate the computing resources at cloud, edge, and terminal layers, and break down the barriers resulted from the diversity and heterogeneity of device owners and manufactures as well as service providers. Second, when designing meta computing technologies, one has to assume that the operating environment is not trust-worthy, or the involved entities have zero-trust among themselves to consider the worst case, and that the hardware and software can be faulty at any time, thus fault-tolerance is a must and the verifiable correctness of the computing results should be emphasized instead of relying on perfect and trustworthy hardware and software for correctness guarantee as traditional computing paradigms do. The requirement of fault-tolerance is mainly driven by the integration of the diverse and heterogeneous resources belonging to different service providers and application domains. Third, to the general public meta computing is a collection of technologies that can virtualize the whole Internet into a computer on-demand, with the underlying resource integration and management transparent to end users, while ensuring that they can get sufficient computing power to satisfy their needs at any time and any location. These three perspectives constitute the kernal of meta computing, which will be further detailed in Section~\ref{sub-sec:design objectives}.


\subsection{Design Objectives} \label{sub-sec:design objectives}

We summarize three major design objectives of meta computing technologies in this subsection.

\noindent {\bf Fault-tolerant computing on the basis of zero trust.}
One of the main innovations of meta computing is the support of fault-tolerant computing based on a zero-trust model. 
Many systems, such as computer, the Internet, a population of living beings, and a society, have undergone an evolution of transitioning from centralized to distributed architectures following a spiral and hierarchical path. Such an evolution not only improves the ability for individuals to survive and collaborate, but also enhances the fault tolerance of the system as a whole. In meta computing, this evolution is essentially a result of trust transfer, i.e., trust is transferred from centralized servers to distributed network entities to form a more trustworthy and easily orchestrating unity. 

Full trust can only be achieved under ideal conditions. For general-purpose scenarios, instead of requiring a 100\% trust, supporting the completion of computing tasks under different trust levels is particularly important, especially when we need to consider resource sharing and consolidation among entities that do not completely trust each other. In an untrusted environment, a strong fault tolerance of computation and storage is particularly beneficial. Fault-tolerance in meta computing requires that the computing outcomes are verifiably correct, even though the computing process and the involved hardware/sofware/data can be faulty. Such a fault-tolerance brings elasticity and flexibility, giving users a stronger confidence to trust the computing services and resource owners a mightier desire to  share their idle computing capabilities. Existing measures can help meta computing reduce errors, while zero-trust fault-tolerance can further tolerate errors. These two types of mechanisms complement each other, making it possible for meta computing to integrate various computing resources hooked on the Internet. 

\noindent {\bf Cloud-edge-terminal resource consolidation making the network a computing power infrastructure.}
Another important objective of meta computing is to integrate the computing resources hooked on the Internet, making use of them to the maximum when they are unused or under-utilized, thereby consolidating the whole network into a computing power infrastructure. In other words, meta computing intends to harvest all possible computing powers scattered all over the Internet and allocate an appropriate amount to a computing task such that the increasingly serious computing power shortage problem can be overcome. To achieve this goal, meta computing needs to break the technical barriers caused by device and protocol heterogeneity as well as the service barriers resulted from different service providers. For the technical barriers, new mechanisms need to be developed to consolidate the resources not only at the network edge but also at the terminal-end, which will be discussed in the next. The service barriers lead to the computing power islands problem, which are mainly caused by service providers who cannot confidently share their resources with each other due to the lack of mutual trust. This issue can be alleviated by the construction of a trusted computing environment among entities that do not trust each other, which has been discussed earlier.

Edge computing focuses on the integration of resources at the network edge, ignoring the huge amount of computing power of the IoT devices at the terminal-end. As IoT devices are extremely resource-constrained with very limited computing capacities, it is hardly possible to transfer the techniques developed for the network edge to integrate the computing resources of a large number of IoT devices. Novel techniques are desperately needed to consolidate their computing powers. It was estimated that there exist about 10 billion active microcontroller-based IoT devices by 2021 and this number will reach 25 billion by 2030, and that the aggregated computing power of a few dozen such devices can be equivalent to that of a personal computer \cite{coremark}. Considering that IoT devices are ubiquitous and close to applications, fully utilizing their computing powers could bring unimaginable benefits. However, the massive scale and severely heterogeneous functional structures of these devices pose extreme challenges for their coherence and consistency. In addition, the data acquisition process requires precise clock synchronization since devices with different data collection cycles, inconsistent data collection results, and asynchronous data transmissions can lead to serious problems, including missing data, inconsistent data, incorrect computation results, uncertain transmission delays, and poor analysis efficiency, which should be addressed by meta computing. 

\noindent {\bf Virtualizing the network into a meta computer.}
The ultimate goal of meta computing is to virtualize the whole network into a meta computer for any task or any person at any time in any where, quickly allocating the right amount of computing power on-demand. In other words, the whole network serves as a giant computing power pool, continuously and elastically outputting computing powers to support various tasks while the underlying resource allocation and management is transparent to the end users, who feel like operating on a easy-to-use computer but have almost no clue regarding how it works, even though the computational and storage resources are distributed around the world and are changing dynamically. 

The design of a meta computer should maximize the ease of use and optimize user experience. For developers, a meta computer should support not message passing but shared memory. The message passing paradigm is suitable for traditional areas such as supercomputers and data centers. However, in applications such as mobile computing, vehicle computing, and the Internet of Things, the interconnection and resource distribution are dynamic, diverse, and complex, and manual optimization for one scenario may worsen the performance of another scenario. Additionally, as the needs to develop distributed applications grow rapidly, it is not appropriate to assume that every developer is an expert in distributed systems and can handle tasks such as performance tuning, system scheduling, and computational fault tolerance, using the message passing paradigm. Therefore, the design of a meta computer should take the shared memory paradigm, in which developers can simply focus on describing the nature of the problem itself as if they were programming for a single machine, while the scheduling of the computational resources, fault tolerance of the computational processes, and details of the computational methods are transparent. The distributed shared memory mechanism provided by the meta computer can be likened to a Just-in-Time (JIT) compiler that takes the abstract computational problem description as input and compiles the application to a concrete computational solution. 
In addition, the meta computer's programming model should include attributes that allow programmers with relevant knowledge or expertise to add directives to the application for advanced optimization, such as the choice of a memory consistency model, manual selection of compute nodes, and different levels of fault tolerance.

\subsection{Architecture}
\begin{figure*}
    \centering
    \includegraphics[width=0.85\textwidth]{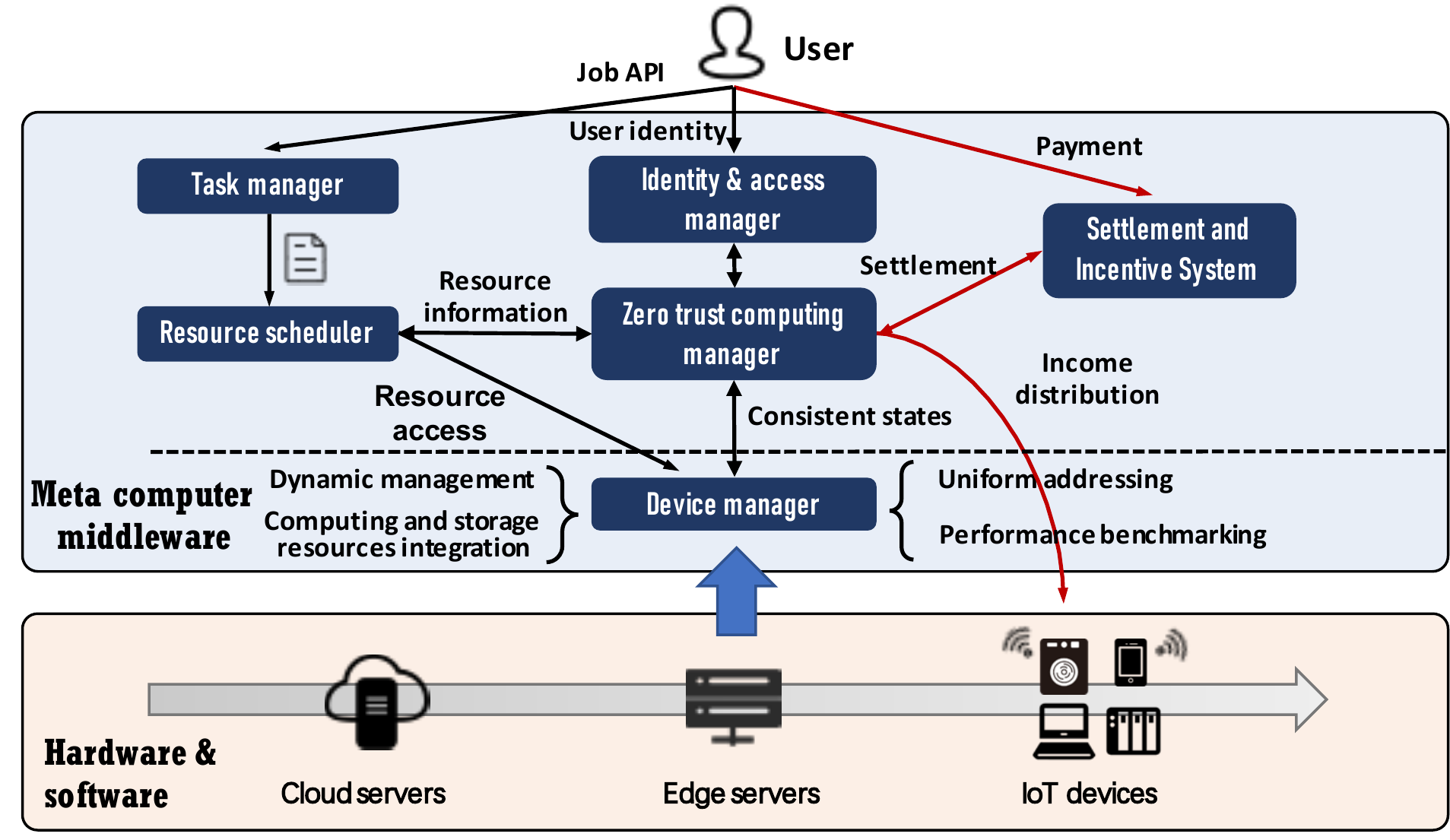}
    \caption{A meta computer architecture}
    \label{fig:architecture}
\end{figure*}

Based on the definition of and perspectives on meta computing, we present an architecture shown in Figure~\ref{fig:architecture}, which consists of the meta computer middleware and the physical hardware \& software. 
A meta computer is middleware-driven, which empowers the meta computing. 
It consists of 
\begin{inparaenum}[(1)]
\item a device management module, 
\item a zero-trust computing module, and
\item a few auxiliary modules. 
\end{inparaenum}
The device management module abstracts the underlying hardware details away, and presents all computing resources in a way that can be easily accessed by the resource scheduler, by reporting the \begin{inparaenum}[(1)]
\item computing capabilities,
\item main memory capabilities,
\item persistent storage capabilities,
\item networking performance and topology, and
\item reliability \& energy efficiency
\end{inparaenum}
of the computing nodes in a unified format.
The underlying hardware details may change over time, hence the device manager module must dynamically report such information to the resource scheduler.
The resource scheduler then dynamically performs online scheduling and optimization of the applications over the network given the resource description.
The zero-trust computing module provides a trustworthy environment for secure and fault-tolerant computing.
In the following we detail the major modules.

\textbf{User}: 
Each user is given a unique digital identity over the network so that they can be identified and verified. 
With that identity, users can submit requests for various computing tasks such as model training, data sharing, and data processing.
The requests only include the task program, input, payment amount, expected output, and other relevant information, and are orthogonal to the hardware configuration details of the meta computer:
\begin{inparaenum}[(1)]
\item the hardware details are time-variant,
\item the mapping of the task to specific hardware resources is handled by the meta computer according to the task specification, and
\item the task programs focus on describing the algorithm rather than the implementation artifacts, so that the meta computer middleware can perform on-line optimizations according to current hardware resource configuration details. 
\end{inparaenum}
To maintain a viable economic model, the users also have private wallet accounts to
\begin{inparaenum}[(1)]
\item pay rents for consuming and
\item receive interests for providing
\end{inparaenum}
computing resources over the network.

\textbf{Identity and Access Manager}: The identity and access manager works with task manager and distributed ledger to share information and ensure secure and fine-grained access control in a capability-based manner.
Capabilities are unforgeable tokens that grant access permissions to certain actions.
They are only accessible to a user or a task when explicitly delegated or granted from another user or task that already has a capability to the resource. 
For example, if user A needs to use user B's data, A must have a legal identity and provide the proper capability granted by B access to that data.

\textbf{Task Manager}: The task manager receives requests from users and preprocesses them.
It decomposes a task based on various constraints (e.g., real-time constraints, resource/budget limits, mission criticality, data security level, etc.), specifying the input and output, data storage location, data boundary, and data flow.
It is also responsible for translating the task so that they could be processed by the resource scheduler.

\textbf{Resource Scheduler}: The resource scheduler is responsible for allocating appropriate resources to each task and computing the resource utilization for final revenue calculation. 
It will
\begin{inparaenum}[(1)]
\item constantly sense the change of the underlying hardware configuration details reported by the device manager,
\item dynamically perform online (re-)compilation and (re-)optimization of submitted tasks and task groups on the fly,
\item dynamically distribute the resulting task artifacts to suitable nodes so that the constraints in the task performance specifications are met and the resource (e.g. memory, power, bandwidth, latency, etc.) utilization is optimized, and
\item constantly ensure that the reliability and security of the tasks conforms to the levels specified.
\end{inparaenum}

\textbf{Device Manager}: The device manager is a crucial component that connects the underlying devices to the rest of the middleware.
Its main purpose is to abstract the resources offered by various heterogeneous devices at the various networking layers into objects or nodes that can be easily accessed by the resource scheduler.
Specifically, it maps computing devices, main memory, persistent storage, etc., of the devices into a Shared Resource Space (SRS), which constitutes a meta computer's hardware configuration. 
When a hardware change in the network is observed, the SRS is dynamically updated and reported to the resource scheduler to facilitate resource scheduling.

\textbf{Zero-trust Computing Manager}: The zero-trust computing manager is responsible for ensuring the consistency of the states via a blockchain. 
Blockchain has established a precedent for creating zero-trust environments and has inspired the developments of other fields~\cite{9743876, 9761745}. 
By utilizing blockchain technologies, a trusted computing environment can be established for users on the basis of zero trust. 
There are two commonly used computing models with blockchain: on-chain computing and off-chain computing. On-chain computing has strong consistency and security, sacrificing the computing efficiency at certain level; on the contrary, off-chain computing reduces the burden of on-chain computing, sacrificing the computing security due to the fact that the corresponding procedure is not directly protected by the chain.
Hence, the choice between the two reflects a trade-off between security and efficiency, and there is a {\em continuum} of mixed approaches between the two extremes for a certain computing task.
To support computation under different trust and security levels, the zero-trust computing manager selects proper computing methods given the locality of trust and establishes a trusted computation environment based on the trust degree of nodes and their required security levels.

\section{Opportunities and Challenges of Meta Computing}\label{sec:opportunities}

\begin{figure*}[htbp]
    \centering
	\includegraphics[width=0.85\textwidth]{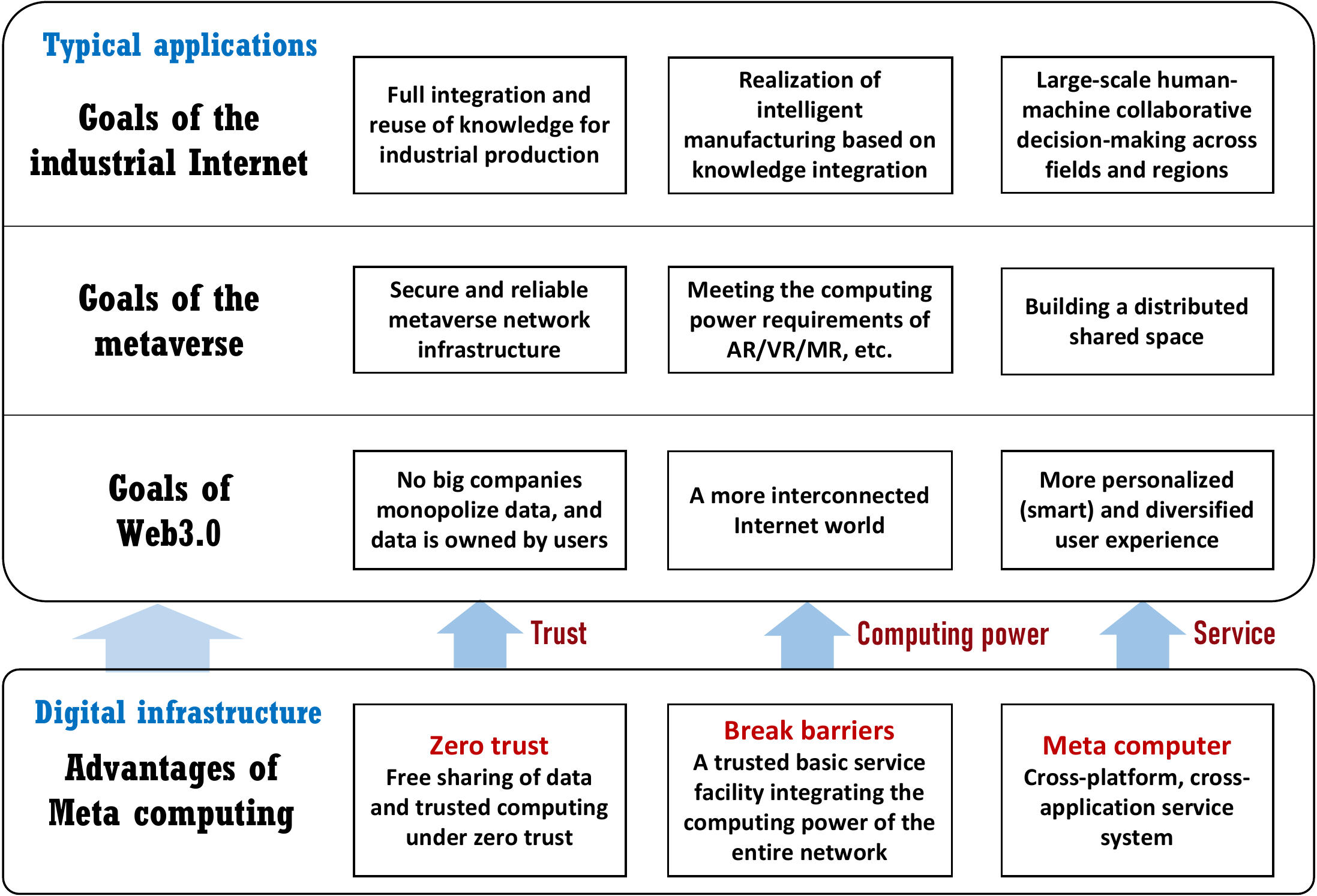}
	\caption{Typical applications of meta computing. }
	\label{fig:applications}
\end{figure*}

\subsection{Opportunities}\label{sub-sec:applications}
In this section, we first suggest a development path for meta computing; then discuss three endogenous applications of meta computing, namely Web 3.0, metaverse, and the Industrial Internet, to demonstrate the main benefits and expected outcomes of meta computing.

\noindent {\bf The Development Path.}
The development of any computing paradigm typically follows a progression from small, local demonstrations to cross-regional and eventually global applications, so does meta computing. Meta computing takes a two-phase incentive to drive the process forward. During the first phase, a meta computer that integrates idle resources within a small area (e.g., a laboratory) is  built to provide better computing services for controlled participants. In the second phase, additional resources from independent organizations or big companies are attracted to join, and the zero-trust computing mechanism helps to provide security and privacy protection, ensuring that the organizations are willing to provide computing powers in exchange for benefits.

\noindent {\bf Web 3.0.}
In the era of Web 3.0, users not only produce and own data, but also disseminate and hold value of the data. Meta computing can support the secure sharing of data in a zero-trust environment, allowing users to choose their own computing resources (e.g., their own devices) and hold the ownership of their data from the moment when the data is generated. Additionally, applications powered by meta computing can use settlement and incentive mechanisms to facilitate the exchange of value through data.

\noindent {\bf Metaverse.}
A Metaverse is a digital space that is open, decentralized, and secure. It is not controlled by any central authority, and individual's data should be protected. Meta computing technologies have several impacts on a Metaverse. First, it provides a safe and reliable network for the Metaverse through Web 3.0. Second,  meta computing can offer the Metaverse enough computing power by integrating network resources and distributing shared memory, which aligns with Metaverse's need for distributed shared space.

\noindent {\bf Industrial Internet.}
The Industrial Internet makes use of the Internet infrastructure to enable the interconnection and data flow between different industrial networks. In this context, the zero-trust environment and incentive mechanism in meta computing can provide higher value returns to knowledge producers, encouraging faster dissemination of knowledge. Meta computing can also provide sufficient computing power to support the distributed collection, storage, training, and learning of data, enabling the intelligence of the Industrial Internet. Meta computers can offer cross-domain services for the industrial Internet, enabling large-scale human-machine collaborative decision-making.

\subsection{Challenges of Meta Computing}\label{sec:challenges}
In this section, we outline the difficulties associated with meta computing from four perspectives.

\noindent {\bf Full resource integration.}
One of the primary challenges of meta computing is to integrate all the resources over the entire network.
Different devices may have different 
\begin{inparaenum}[(1)]
\item computing capability,
\item main memory capacity,
\item storage capacity and
\item communication quality, let alone their different
\item reliability and
\item energy efficiency.
\end{inparaenum}
Thus, the challenge to run a single application over the network boils down to how to 
\begin{inparaenum}[(1)]
\item distribute the entire application onto these resources according to its Quality-of-Service (QoS) specification while 
\item minimizing the overall resource usage.
\end{inparaenum}

When multiple applications that have different criticality and budget need to share all these computing resources over the network, an even more challenging combinatorial scheduling problem presents itself: \begin{inparaenum}[(1)]
\item when the applications can fit into the resources, a scheme that accommodates all applications needs to be generated, and
\item when the resources cannot meet all the requirements (which is the common case), we need to decide what application should run with what resources, and what application to drop.
\end{inparaenum}
Additionally, in the case where an application’s QoS can be partially met, the scheme must be able to gracefully degrade its service quality.

In the discussion above, we’ve made an assumption that the
\begin{inparaenum}[(1)]
\item computing resources,
\item application set, and
\item QoS requirements
\end{inparaenum}
are known priori and time-invariant.
However, in practice, all of the aspects mentioned above are dynamic, which forces to seek an online algorithm. The computing resources may only intermittently exist due to transient faults or scheduled maintenance, and applications may be dynamically booted or halted. Even for a single application, its QoS requirements may be different at different stages of execution.

To conclude, the difficulties in integrating resources remain a challenge that needs to be seriously addressed.

\noindent {\bf Zero-trust computing.}
Another big challenge is how to collaborate on a computing task over a network where potentially no or very little trust exists.
Prior computing paradigms, i.e. cloud-computing and grid-computing, assumes mutual trust between the involved parties.
Hence, these computing paradigms only need to handle transient faults which are random in nature, rather than deliberate security breaches and miscalculations that they cannot handle.

An emerging technology, the blockchain, addresses the lack of mutual trust with consensus algorithms such as the Proof-of-Work (PoW)~\cite{nakamoto2008bitcoin}:
\begin{inparaenum}[(1)]
\item all nodes spend their computational resources on a single difficult computational task, and 
\item when completed, competitively broadcasts their results over the entire network, so the states of all the nodes are synchronized.
\end{inparaenum}
However, in this mechanism,
\begin{inparaenum}[(1)] 
\item all nodes perform essentially competition on the same task rather than collaborate on different portions of the task,
\item the chosen task is usually artificially constructed and the computation power is simply wasted, and \item the paradigm does not allow multitasking. 
\end{inparaenum}
Hence, this paradigm does not suit the meta-computing well.

In meta-computing, one needs to
\begin{inparaenum}[(1)] 
\item assume zero-trust between random pairs of nodes over the entire network,
\item dynamically measure the mutual trust between the nodes so as to expose the locality-of-trust,
\item organize the nodes according to such locality in the hope that the nodes collaborate on tasks as much as possible while leaving sufficient mutual competition for check-and-balance,
\item distribute the tasks over such node organization taking the computing resources into account, and
\item be able to mutually isolate different tasks spatially and temporally with the exception of information sharing in a controlled manner.
\end{inparaenum}
To this end, this computing paradigm remains a challenge despite prior research.

\noindent {\bf Programming paradigm.}
For meta-computing, the programming paradigm is another challenge. 
Existing programming paradigms either assume
\begin{inparaenum}[(1)] 
\item a single {\em synchronous} computing machine programming model where all computation resources are local, or
\item a vast number of {\em asynchronous} computing nodes that communicate with each other over the network.
\end{inparaenum}
They require the programmer to write the application in an topology-aware fashion, as the programmer assumes full control of the communication within or between the nodes, either synchronous (plain shared memory accesses) or asynchronous (remote send/receive).
In the latter case, the programmer often needs to manually implement fault-tolerant features in case the nodes inadvertently go offline, which adds to the complexity of writing the applications. 
 
In meta computing, however, the underlying computing infrastructure is not predetermined and may be a continuum between the two extremes.
If the application is written in a way that is explicitly optimized for one computing topology, its performance may be lackluster on other topologies due to mismatch between the optimization assumptions and the underlying hardware.
Moreover, such optimizations often contain many details that obfuscate the algorithm, so porting to other topologies require a manual rewrite or re-optimization of the whole program. 
Hence, when programming a meta computer, we cannot assume the underlying hardware topology; even if the topology is known at a given timepoint, it may change dynamically in the future. 
The programming paradigm therefore should describe the algorithm of computation task (``{\em what}'') instead of its implementation details (``{\em how}'').
The mapping from the algorithm to the implementation detail should be undertaken by a meta compiler that {\em dynamically} performs the profiling, optimization and re-optimization during the execution of the program to cope with the changes in computing topology.
To this end, the programming paradigm should fully expose the mathematical components of the algorithm so that the compiler can optimize the algorithm as much as possible.

Moreover, the programming paradigm for a meta computer needs to be compatible with existing programming paradigms and programming languages, which means that the existing applications could be converted to run on the meta computer, albeit at a performance penalty.
It is also desirable that the paradigm could assume a Distributed Shared Memory space (DSM) or Distributed Shared Object Space (DSOS) pattern to simplify programming for average programmers that are already familiar with single-machine programming.
At last, the programming paradigm also needs to provide sufficient means so that the different programs can interact with each other in a controlled manner.

\noindent {\bf Security and privacy.}
Finally, security and privacy is always a concern for all computing paradigms, and meta computing is no exception to this.
Computer system security and privacy largely relies on the notion of {\em protection domains}: a protection domain
\begin{inparaenum}[(1)]
\item {\em isolates} an application from other applications so that other application cannot compromise its security, and
\item {\em confines} an application to within itself so that it cannot compromise or attack other applications.
\end{inparaenum}
In a single machine, the efficient protection domains are usually implemented with dedicated hardware such as Memory Management Units (MMUs) or virtualization extensions, and the implementations are largely hardware-specific. 
Applications are booted up as processes that are isolated from each other, and all applications alongside with their data are located on the local machine.
However, in a meta computer, an application may {\em span} multiple physical nodes distributed over the network, and the protection domain that contains the application also needs to span multiple nodes to protect the application.
Different nodes may have different protection domain hardware idiosyncrasies, while providing different levels of {\em isolation} and {\em confinement} quality. A unified protection domain description scheme is needed to
\begin{inparaenum}[(1)]
\item abstract the hardware implementation details away and
\item provide the same protection strength over the different platforms.
\end{inparaenum}

Moreover, it might also be necessary to expand existing behavioral monitoring, identity authentication and intrusion detection technologies over multiple physical nodes when there is no fixed mapping between physical nodes and applications or users.

\section{Conclusions}\label{conclusion}
In this paper, we propose meta computing, a novel computing paradigm that intends to turn the whole Internet and its hooked resources into a giant computing power pool to enable the emerging data-intensive applications with strong realtime and security requirements. Meta computing opens a new arena for academia and industry to invent novel technologies to handle the challenges brought by the zero-trust or low-trust computing environments as well as the error-prone and heterogeneous low-end ubiquitous terminal devices. Facilitated by meta computing technologies, the Internet can be virtualized into a meta computer with sufficient computing power for any task or any person while the underlying on-demand resource allocation and trust management are transparent to end users. We propose a meta computer architecture, analyze its design objectives and challenges, and identify a few endogenous applications. Meta computing is deemed as an innovative solution for the increasingly serious computing power shortage problem. 


\bibliographystyle{IEEEtran}
\bibliography{reference}

\begin{thebibliography}{10}
\providecommand{\url}[1]{#1}
\csname url@samestyle\endcsname
\providecommand{\newblock}{\relax}
\providecommand{\bibinfo}[2]{#2}
\providecommand{\BIBentrySTDinterwordspacing}{\spaceskip=0pt\relax}
\providecommand{\BIBentryALTinterwordstretchfactor}{4}
\providecommand{\BIBentryALTinterwordspacing}{\spaceskip=\fontdimen2\font plus
\BIBentryALTinterwordstretchfactor\fontdimen3\font minus
  \fontdimen4\font\relax}
\providecommand{\BIBforeignlanguage}[2]{{%
\expandafter\ifx\csname l@#1\endcsname\relax
\typeout{** WARNING: IEEEtran.bst: No hyphenation pattern has been}%
\typeout{** loaded for the language `#1'. Using the pattern for}%
\typeout{** the default language instead.}%
\else
\language=\csname l@#1\endcsname
\fi
#2}}
\providecommand{\BIBdecl}{\relax}
\BIBdecl

\bibitem{chellappa1997intermediaries}
R.~Chellappa, ``Intermediaries in cloud-computing: A new computing paradigm,''
  in \emph{INFORMS Annual Meeting, Dallas}, 1997, pp. 26--29.

\bibitem{oram2001peer-to-peer}
A.~Oram, \emph{Peer-to-Peer: Harnessing the power of disruptive
  technologies}.\hskip 1em plus 0.5em minus 0.4em\relax " O'Reilly Media,
  Inc.", 2001.

\bibitem{barkai2001technologies}
D.~Barkai, ``Technologies for sharing and collaborating on the net,'' in
  \emph{Proceedings First International Conference on Peer-to-Peer
  Computing}.\hskip 1em plus 0.5em minus 0.4em\relax IEEE, 2001, pp. 13--28.

\bibitem{itu2005internet}
I.~ITU, ``Internet reports 2005: The internet of things,'' \emph{Geneva: ITU},
  2005.

\bibitem{satyanarayanan2009case}
M.~Satyanarayanan, P.~Bahl, R.~Caceres, and N.~Davies, ``The case for vm-based
  cloudlets in mobile computing,'' \emph{IEEE pervasive Computing}, vol.~8,
  no.~4, pp. 14--23, 2009.

\bibitem{bonomi2012fog}
F.~Bonomi, R.~Milito, J.~Zhu, and S.~Addepalli, ``Fog computing and its role in
  the internet of things,'' in \emph{Proceedings of the first edition of the
  MCC workshop on Mobile cloud computing}, 2012, pp. 13--16.

\bibitem{coremark}
``{CoreMark: An EEMBC Benchmark, https://www.eembc.org/coremark/scores.php,
  retrieved 02/03/23},'' 2023.

\bibitem{witt1965ibm}
B.~Witt and L.~Ward, ``Ibm operating system/360 concepts and facilities,''
  \emph{IBM Systems Reference Library}, pp. 1--91, 1965.

\bibitem{Foster1998grid}
I.~T. Foster and C.~Kesselman, \emph{The Grid, Blueprint for a New Computing
  Infrastructure}.\hskip 1em plus 0.5em minus 0.4em\relax Morgan Kaufmann
  Publishers, 1998.

\bibitem{chasins2022sky}
S.~Chasins, A.~Cheung, N.~Crooks, A.~Ghodsi, K.~Goldberg, J.~E. Gonzalez, J.~M.
  Hellerstein, M.~I. Jordan, A.~D. Joseph, M.~W. Mahoney \emph{et~al.}, ``The
  sky above the clouds,'' \emph{arXiv preprint arXiv:2205.07147}, 2022.

\bibitem{krol2019compute}
M.~Kr{\'o}l, S.~Mastorakis, D.~Oran, and D.~Kutscher, ``Compute first
  networking: Distributed computing meets icn,'' in \emph{Proceedings of the
  6th ACM Conference on Information-Centric Networking}, 2019, pp. 67--77.

\bibitem{weiser1999computer}
M.~Weiser, ``The computer for the 21st century,'' \emph{ACM SIGMOBILE mobile
  computing and communications review}, vol.~3, no.~3, pp. 3--11, 1999.

\bibitem{zheng2003ubiquitous}
Z.~Zheng and Z.~Wu, ``A survey of pervasive computing,'' \emph{Computer
  Science}, vol.~30, pp. 18--29, 2003.

\bibitem{sun2010internet}
Q.~Sun, J.~Liu, S.~Li, C.~Fan, and J.~Sun, ``Internet of things: Summarize on
  concepts, architecture and key technology problem,'' \emph{Journal of Beijing
  University of Posts and Telecommunications}, vol.~33, no.~3, p.~9, 2010.

\bibitem{itu2009overview}
Y.~ITU-T, ``Overview of ubiquitous networking and of its support in ngn,''
  \emph{ITU-T Recommendation}, 2009.

\bibitem{wolf2009cyber}
W.~Wolf, ``Cyber-physical systems,'' \emph{Computer}, vol.~42, no.~03, pp.
  88--89, 2009.

\bibitem{dilley2002globally}
J.~Dilley, B.~Maggs, J.~Parikh, H.~Prokop, R.~Sitaraman, and B.~Weihl,
  ``Globally distributed content delivery,'' \emph{IEEE Internet Computing},
  vol.~6, no.~5, pp. 50--58, 2002.

\bibitem{warneke2001smart}
B.~Warneke, M.~Last, B.~Liebowitz, and K.~S. Pister, ``Smart dust:
  Communicating with a cubic-millimeter computer,'' \emph{Computer}, vol.~34,
  no.~1, pp. 44--51, 2001.

\bibitem{9743876}
M.~Xu, F.~Zhao, Y.~Zou, C.~Liu, X.~Cheng, and F.~Dressler, ``Blown: A
  blockchain protocol for single-hop wireless networks under adversarial
  sinr,'' \emph{IEEE Transactions on Mobile Computing}, pp. 1--1, 2022.

\bibitem{9761745}
M.~Xu, Z.~Zou, Y.~Cheng, Q.~Hu, D.~Yu, and X.~Cheng, ``Spdl: A
  blockchain-enabled secure and privacy-preserving decentralized learning
  system,'' \emph{IEEE Transactions on Computers}, vol.~72, no.~2, pp.
  548--558, 2023.

\bibitem{nakamoto2008bitcoin}
S.~Nakamoto, ``Bitcoin: A peer-to-peer electronic cash system,''
  \emph{Decentralized business review}, p. 21260, 2008.

\end{thebibliography}

\end{document}